\documentclass[%
 aip,
 amsmath,amssymb,
preprint,%
]{revtex4-1}

\usepackage{graphicx}
\usepackage{dcolumn}
\usepackage{bm}

\usepackage[utf8]{inputenc}
\usepackage[T1]{fontenc}
\usepackage{mathptmx}
\usepackage{etoolbox}

\usepackage[colorlinks=true,linkcolor=blue]{hyperref}

\makeatletter
\def\@email#1#2{%
 \endgroup
 \patchcmd{\titleblock@produce}
  {\frontmatter@RRAPformat}
  {\frontmatter@RRAPformat{\produce@RRAP{*#1\href{mailto:#2}{#2}}}\frontmatter@RRAPformat}
  {}{}
}%
\makeatother
\begin{document}


\title[]{Next-Generation Time-Resolved Scanning Probe Microscopy
\\ - Ushering in a New Era of Nanoscale Exploration with Enhanced Usability}
\author{Katsuya Iwaya}
\affiliation{%
UNISOKU Co., Ltd., Hirakata, Osaka 573-0131, Japan
}%

\author{Hiroyuki Mogi}
\affiliation{ 
Faculty of pure and applied sciences, University of Tsukuba, Tsukuba, Ibaraki 305-8573, Japan
}%

\author{Shoji Yoshida}
\affiliation{ 
Faculty of pure and applied sciences, University of Tsukuba, Tsukuba, Ibaraki 305-8573, Japan
}%

\author{Yusuke Arashida}
\affiliation{ 
Faculty of pure and applied sciences, University of Tsukuba, Tsukuba, Ibaraki 305-8573, Japan
}%

\author{Osamu Takeuchi}
\affiliation{ 
Faculty of pure and applied sciences, University of Tsukuba, Tsukuba, Ibaraki 305-8573, Japan
}%

\author{Hidemi Shigekawa*}
\email{hidemi@ims.tsukuba.ac.jp}
\affiliation{ 
Faculty of pure and applied sciences, University of Tsukuba, Tsukuba, Ibaraki 305-8573, Japan
}%

\date{\today}

\begin{abstract}
Understanding the nanoscale carrier
dynamics induced by light excitation
is the key to unlocking futuristic
devices and innovative functionalities in
advanced materials. Optical pump-probe
scanning tunneling microscopy (OPP-STM)
has opened a window to these phenomena.
However, mastering the combination of
ultrafast pulsed lasers with STM requires
high expertise and effort. We have shattered
this barrier and developed a compact
OPP-STM system accessible to all.
This system precisely controls laser pulse
timing electrically and enables stable laser
irradiation on sample surfaces. Furthermore,
by applying this technique to atomic
force microscopy (AFM), we have captured
time-resolved force signals with an
exceptionally high signal-to-noise ratio.
Originating from the dipole-dipole interactions,
these signals provide insights into
the carrier dynamics on sample surfaces,
which are activated by photo-illumination.
These technologies are promising as powerful
tools for exploring a wide range of
photoinduced phenomena in conductive
and insulating materials.
\end{abstract}

\maketitle

\section{Introduction}

The need for exceptional spatial and temporal
resolutions is at the heart of deciphering
the intricate carrier dynamics in nanoscale
materials. Conventional scanning tunneling
microscopy (STM), although providing
unparalleled spatial and energy resolutions,
hits a temporal resolution ceiling at
sub-millisecond levels owing to preamplifier
bandwidth constraints. With the integration
of transformative OPP techniques into STM,
such barriers have been overcome, realizing
higher temporal resolutions. This fusion is
pivotal in probing materials' complex nonequilibrium
carrier and spin dynamics\cite{Terada,Yoshida_NatNanotech}.
Another technique, the use of a subcycle
electric field as bias voltage, has been incorporated
as an electric-field-driven STM.
This technique provides temporal resolutions
of less than 1 ps and 30 fs, maintaining
the spatial resolution of STM using terahertz
(THz) and mid-infrared pulses\cite{Cocker, Garg, Yoshida_ACSPhoton, Arashida_ACSPhoton}.
This leap in technology widens the horizon
of time-resolved STM; however, it still requires
high expertise and simplification of
the system to make it accessible for a broad
spectrum of research endeavors. In addition,
since the application of STM is limited to
conducting materials, extending this OPP
technique to atomic force microscopy (AFM)
would further expand the capabilities of this
time-resolved measurement technique.

\section{Optical Pump-Probe Scanning Tunneling
Microscopy (OPP-STM)}

In the conventional OPP method, pump light
and probe light, which are delayed in time,
are irradiated onto a sample, as shown in
Figure~1a. When carriers such as electrons
and holes excited by the pump light remain
in an excited state, the excitation by the
probe light is suppressed (absorption bleaching).
Therefore, by measuring the reflectivity
of the probe light as a function of the delay
time, one can investigate the dynamics of
the states excited by the pump light with a
time resolution corresponding to the pulse
width of the excitation light. However, the
spatial resolution is limited to the diffraction
and light-spot size (around micrometer
order).

In OPP-STM, the sample surface under the
STM tip is first excited by a pump pulse and
subsequently by a probe pulse with a delay
time $t_{\rm d}$, and the tunneling current is detected
using a conventional preamplifier (Fig.~1b)
\cite{Terada}. When $t_{\rm d}$ is sufficiently long ((1) in Fig.~1c),
most of the photocarriers excited by the pump
pulse relax to the ground state before the
subsequent probe pulse illumination so that
a similar number of carriers would be excited
by the probe pulse as with the pump pulse,
resulting in a large transient current $I^*_{\rm probe}$. In
contrast, when $t_{\rm d}$ is short, the excited states
remain occupied with the photocarriers excited
by the pump pulse when the probe pulse
illuminates the sample so that the optical absorption
saturates, resulting in a small $I^*_{\rm probe}$
((3) in Fig.~1c). By illuminating a pair of pump
and probe pulses sequentially and by varying
$t_{\rm d}$, one can detect a time-averaged tunneling
current $<I>$ as a function of $t_{\rm d}$ (Fig.~1d). By
fitting the time-resolved tunneling current
with an exponential function, we can obtain
a decay time at the tip location, as examples
are shown in Figure~1e\cite{Terada}.

\begin{figure*}
\includegraphics[width=15cm]{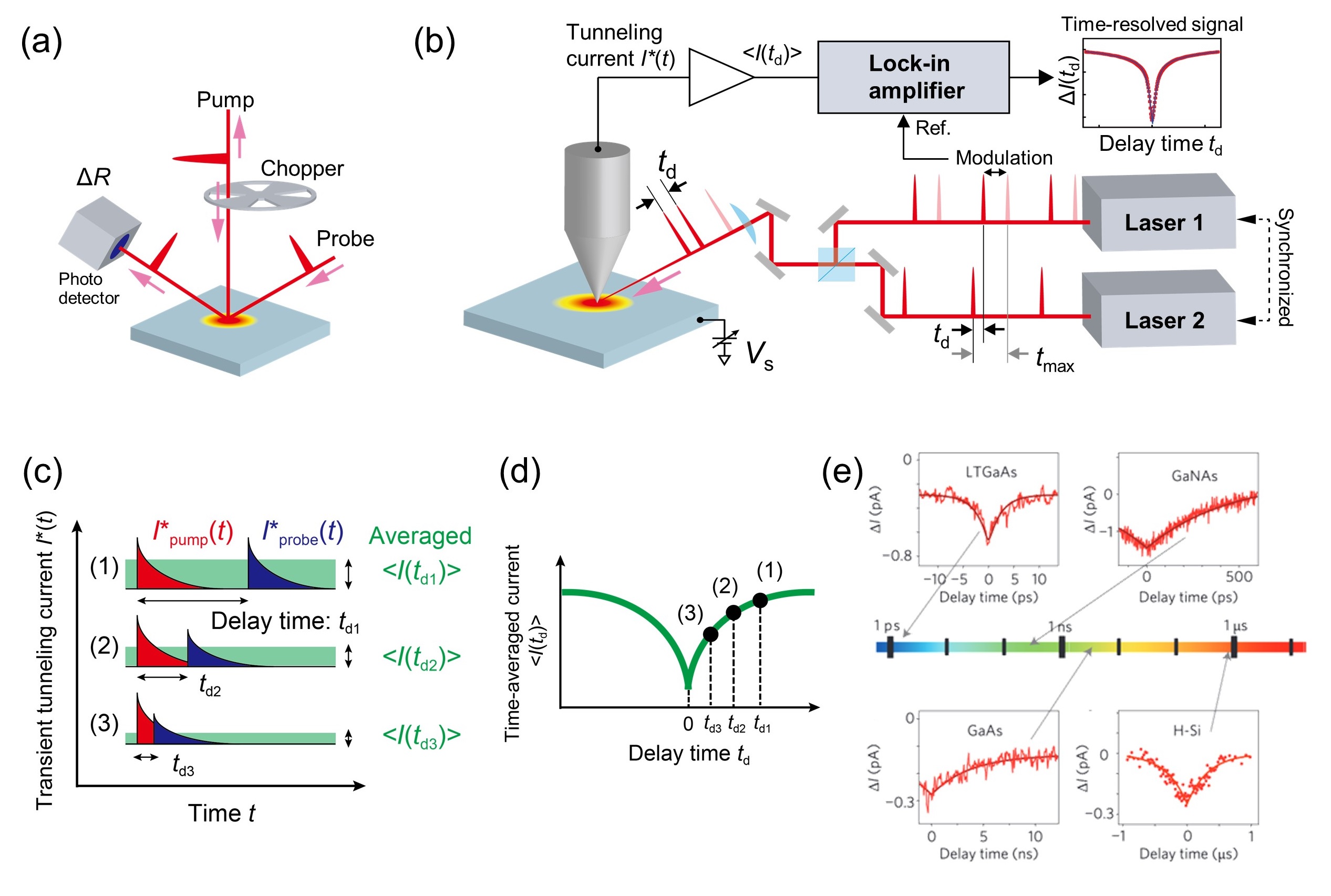}
\caption{(a) Conventional OPP configuration. 
(b) Schematic of OPP-STM with the delay-time
modulation method. 
(c) Transient tunneling current $I^*$ induced by the pump light and probe light
at the representative delay time $t_{\rm d}$. Time-averaged tunneling current $<I(t_{\rm d})>$ is shown for each case (green rectangular). 
(d) $<I>$ as a function of $t_{\rm d}$. The time-averaged $<I>$ corresponding to each case in (c) is plotted. (e) Time-resolved signals obtained for several samples\cite{Terada}.
}
\end{figure*}

In the macroscopic OPP technique, the
modulation of optical intensity is conventionally
utilized to detect a weak OPP-induced
signal. However, the optical intensity
modulation causes severe problems, such as
the thermal expansion of the STM tip. Since
changes in tip-sample distance are exponentially
multiplied in the tunneling current,
the optical intensity modulation cannot be
directly applied to STM. To suppress the thermal
expansion effect, we have developed an
excellent delay-time modulation technique\cite{Terada}. 
In this technique, we use two delay times
($t_{\rm d}$ and $t_{\rm max}$ in Fig.~1b). 
The longer delay time
$t_{\rm max}$ is generally set to one-half of the laser
pulse interval (for example, 0.5~$\mu {\rm s}$ for 1~MHz
repetition rate), corresponding to the longest
delay time available for the selected
repetition rate. We modulate the delay time
between $t_{\rm d}$ and $t_{\rm max}$ at, for example, 1~kHz
and detect the resultant tunneling current
$\Delta I(t_{\rm d}) = <I(t_{\rm d})>$ $-$ $<I(t_{\rm max})>$ 
using the lock-in amplifier (Fig.~1b). 
This modulation technique enables us to keep the thermal load at
the tunnel junction constant, substantially
suppressing the thermal expansion effect.

\begin{figure*}
\includegraphics[width=16cm]{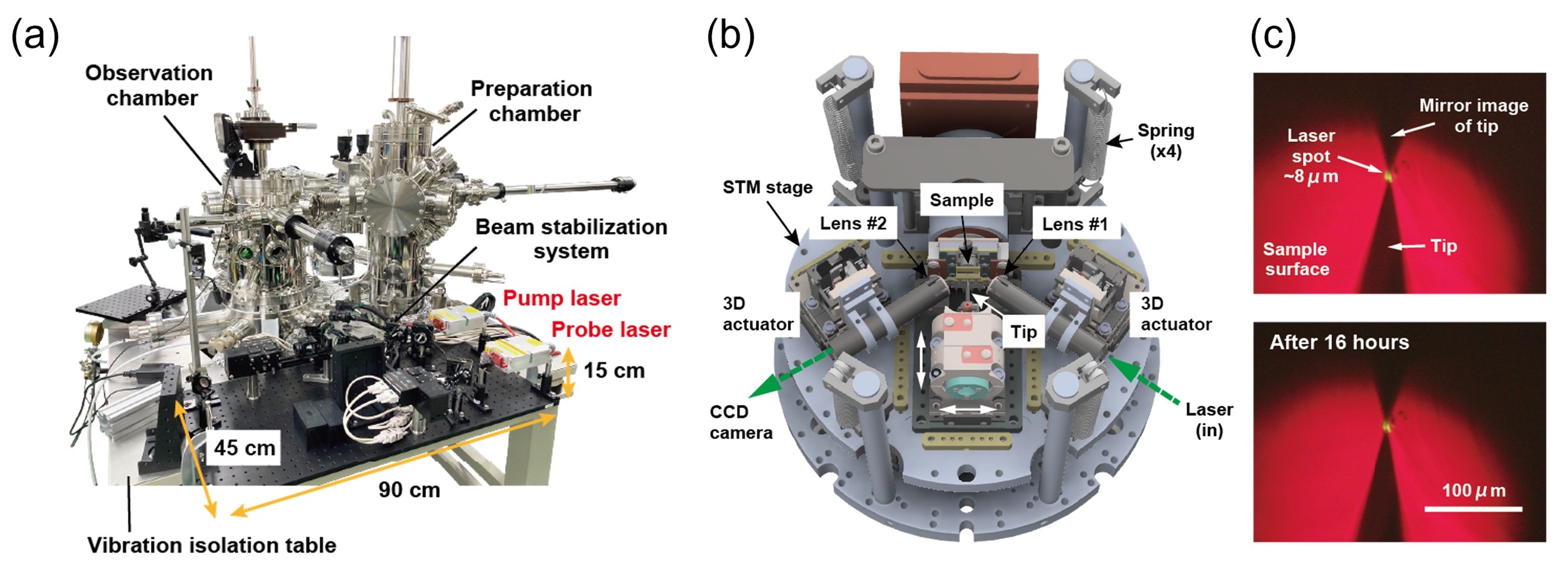}
\caption{
(a) Photograph of the compact OPP-STM system. (b) Three-dimensional illustration of the OPP-STM unit. (c) Optical image of the tip and
its mirror image on GaAs(110) surface with a laser spot illuminated at the tunneling junction before (top) and after 16 h (bottom), showing the
stability of the laser spot\cite{Iwaya}.
}
\end{figure*}

\section{Compact and Stable OPP-STM}

The delay time modulation technique has enabled
reliable OPP-STM measurements. Since
then, many studies such as the atomic scale
carrier dynamics around a single impurity on
a GaAs surface\cite{Yoshida_APEX, Kloth} and the visualization of
the ultrafast carrier dynamics in a GaAs-PIN
junction\cite{Yoshida_Nanoscale} have been reported. 
However, the optical system has been generally complex
and large in scale, hindering the widespread
use of this technique. To overcome this difficulty,
the OPP-STM system, whose laser-pulse
timing is electrically controlled by external
triggers, has significantly improved the ease
of use, but its temporal resolution has been
limited to the nanosecond range\cite{Mogi_APEX1}. 
In addition, fluctuations in optical intensity cause
unexpected issues, such as the thermal expansion
effect, making the accurate observation
of physical phenomena challenging.

To improve the temporal resolution and the
stability of laser illumination based on a
compact electrically controlled laser system
(Fig.~2a), the OPP-STM system with a temporal
resolution of tens of picoseconds has
recently been developed\cite{Iwaya}. The long-term
stability of laser illumination was realized
also by placing the focus lens on the STM
stage (Fig.~2b) and confirmed by monitoring
the tip and its mirror image on the sample
surface together with a laser spot focused on
the tip-sample junction (Fig.~2c).

\begin{figure*}
\includegraphics[width=15cm]{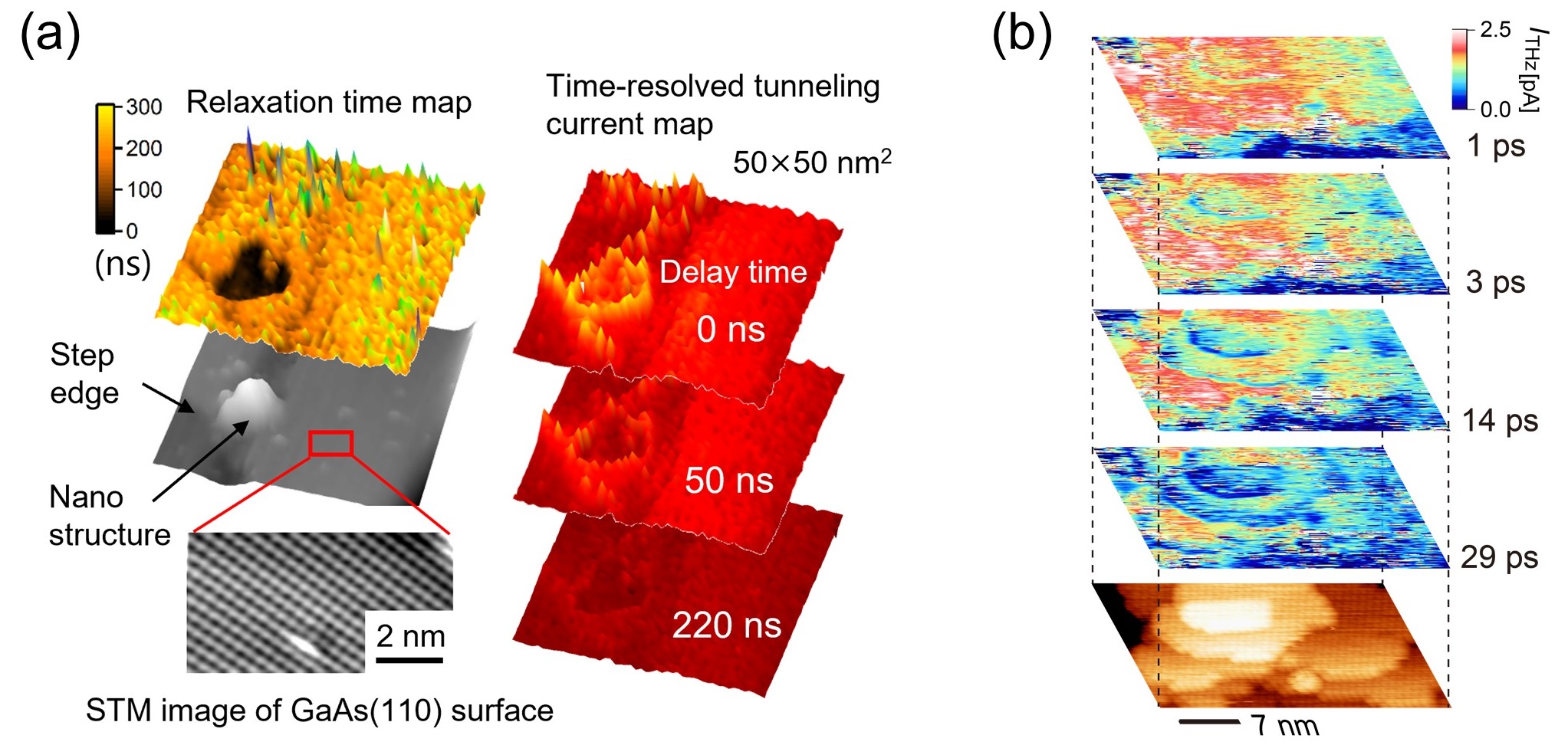}
\caption{
(a) Grid-point time-resolved tunneling current measurement on GaAs(110) cleaved surface
\cite{Iwaya}. Characteristic features along the step edge and the perimeter of the nanoscale bump
structure are identified in the time-resolved tunneling current map (right). By fitting a time-resolved
tunneling current curve at each grid point, we can obtain the nanoscale relaxation time
map (top left). (b) Snapshots of ultrafast motion of photoinjected electrons in a C$_{60}$ multilayer/
Au sample obtained by THz-STM\cite{Yoshida_ACSPhoton}.
}
\end{figure*}

To demonstrate the performance of the
system, we conducted a grid-point time-resolved
tunneling current measurement on
GaAs(110) surfaces at $T = 6$~K (Fig.~3a). By
measuring the time-resolved tunneling current
at each grid point, we can compile the
time-resolved tunneling current maps at each
delay time and determine characteristic features
along the nanostructure (Fig.~3a, right).
Furthermore, it is possible to map a relaxation
time by fitting each curve with an exponential
function. The relaxation time map (Fig.~3a, top left) demonstrates that the relaxation time inside the nanoscale bump structure
is substantially shorter than that of its surroundings.
This mapping technique is powerful
in visualizing carrier dynamics associated
with nanoscale structures. Figure~3b
shows an example of a time-resolved signal
map obtained by THz-STM. The improvement
in the usability of the optical system can be
extended to electric-field-driven STM, which
is currently our ongoing investigation.

\section{OPP Atomic Force Microscopy}

The optical system we have developed is also
applicable to AFM, expanding its potential
beyond conductive materials such as semiconductors
and metals, traditionally targeted
in STM. Integrating AFM into our repertoire
significantly broadens the horizons of nanoscale
time-resolved microscopic techniques.
Previous studies have clarified the
photoexcited dynamics of forces originating
from surface photovoltage and dipole-dipole
interactions (Fig.~4a) based on optical intensity
modulation (Fig.~1a)\cite{Jahng_APL, Schumacher_APL}. We have
advanced these achievements by combining
electrically controlled delay-time modulation
with tuning-fork-type frequency modulation
(FM) AFM (Fig.~4b)\cite{Mogi_APEX_AFM}. 
By focusing
optical pulse pairs onto the apex of the AFM
probe, we were able to detect the amplitude
of the frequency shift $\Delta f$ as a time-resolved
signal. The markedly stable time-resolved
measurements on bulk WSe$_2$, a layered semiconductor,
have unveiled dynamics with
two decay components through an ultrafast
photoinduced force (Fig.~4c). These signals,
decoded as the surface recombination and
diffusion of photocarriers through tunneling
current and force spectroscopy, mark a significant
stride in our understanding of material
properties. This innovative approach not
only resolves the limitations encountered in
time-resolved STM due to tunneling current
but also paves the way for its application
across a diverse array of materials, setting a
new standard in nanoscale imaging.

\begin{figure*}
\includegraphics[width=15cm]{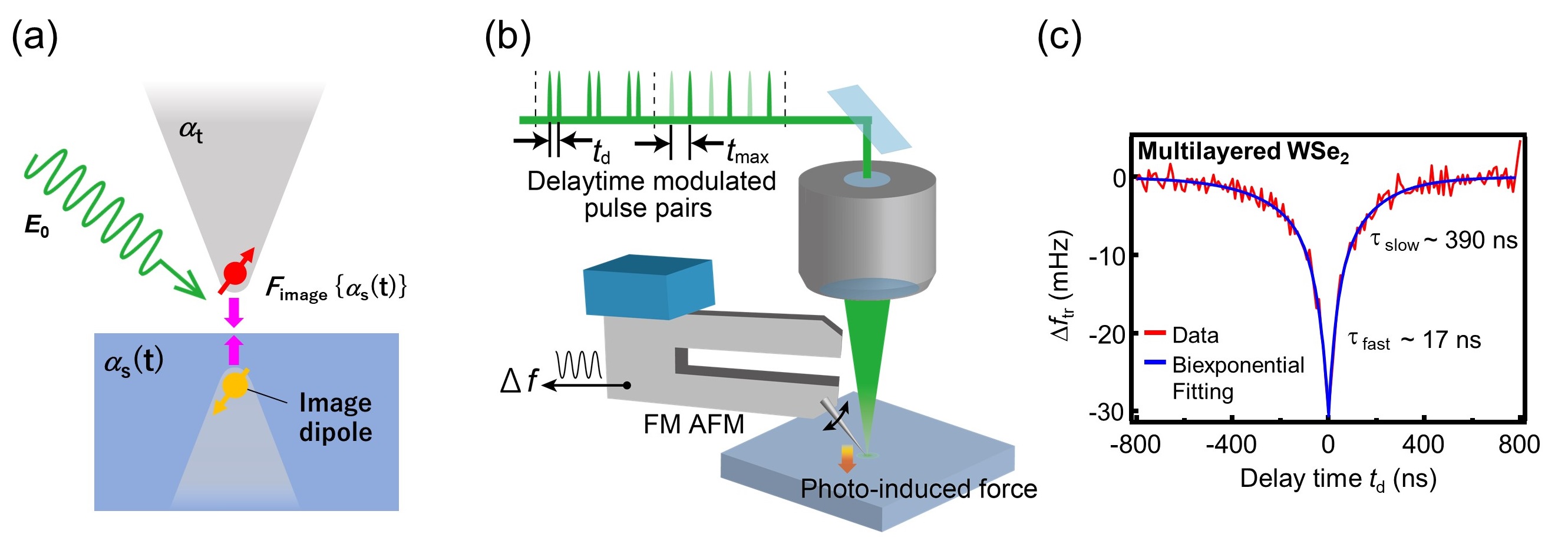}
\caption{
(a) Dipole-dipole interaction mechanism. $\alpha_{\rm t}$, $\alpha_{\rm s}$, and $E_{\rm 0}$ are the complex effective polarizabilities
of the tip and sample, the incident light electric field vector, respectively. (b) Schematic
of tuning-fork-type time-resolved FM AFM setup. (c) Time-resolved signal obtained for a multilayer
WSe$_2$ sample and fitting results\cite{Mogi_APEX_AFM}.
}
\end{figure*}

\section{Conclusion}
Our streamlined optical system markedly eases
researchers' path to harness the transformative
power of the OPP-STM technique. With
the multiprobe (MP) system, features such as
small islands on an insulating substrate can
be observed by using one tip as an electrode
while conducting STM measurements with
the other tip. The OPP-MP measurements
have already been demonstrated on monolayer
transition metal dichalcogenides to investigate
nanoscale exciton dynamics\cite{Mogi_NPJ, Mogi_JJAP}.
Regarding the limitations of wavelength and
temporal resolutions, cutting-edge laser technology,
which has been rapidly developing,
may overcome these limitations, thereby realizing
a higher-performance optical system
enabling a wavelength-variable, externally
controllable laser system with a smaller pulse
width in the future. 
The development of easy-to-use optical systems and their applications
to various scanning probe techniques will expand
the capability of OPP-SPM techniques
and contribute to a deeper understanding of
various photo-induced phenomena.


\end{document}